\documentclass[aps,twocolumn]{revtex4-1}

\usepackage{graphicx,dcolumn,bm,comment,color,braket}

\begin{document}

\preprint{APS/123-QED}

\title{Chaos on a High-Dimensional Torus}

\author{Jumpei F. Yamagishi}
\author{Kunihiko Kaneko}
 \email{kaneko@complex.c.u-tokyo.ac.jp}
 \affiliation{Graduate School of Arts and Sciences, The University of Tokyo, 3-8-1 Komaba, Meguro-ku, Tokyo 153-8902, Japan}

\date{\today}

\begin{abstract}
Transition from quasiperiodicity with many frequencies (i.e., a high-dimensional torus) to chaos is studied by using $N$-dimensional globally coupled circle maps. First, the existence of $N$-dimensional tori with $N\geq 2$ is confirmed while they become exponentially rare with $N$. Besides, chaos exists even when the map is invertible, and such chaos has more null Lyapunov exponents as $N$ increases. This unusual form of ``chaos on a torus,'' termed toric chaos, exhibits delocalization and slow dynamics of the first Lyapunov vector. Fractalization of tori at the transition to chaos is also suggested. The relevance of toric chaos to neural dynamics and turbulence is discussed in relation to chaotic itinerancy. 
\end{abstract}

\maketitle

Quasiperiodic motion with multiple incommensurate frequencies is 
ubiquitously observed in nature
\cite{Kuramoto,Berge-Pomeau-Vidal,KK-thesis,Ott,Vulpiani} 
including hydrodynamics \cite{hydrodynamics,Experiment34}, semiconductor lasers \cite{semiconductor}, electric circuits \cite{ElectricCircuit}, chemical reactions \cite{Chemical}, heart rhythm \cite{cardiac}, and the brain \cite{brain}; while bifurcation(s) from quasiperiodicity to chaos appears in general as a parameter of the system in concern is changed.

This transition from periodic or quasiperiodic motions to chaos gathered much attention in the 1970s and 80s, where several routes to chaos were unveiled in theory, simulations, and experiments \cite{Eckmann1981,Berge-Pomeau-Vidal,KK-thesis,Ott,Vulpiani}. They include a period-doubling route to chaos with its renormalization-group analysis \cite{Feigenbaum1979} and intermittency \cite{Intermittency}, as well as the transition from quasiperiodicity (on a torus attractor) to chaos \cite{Feigenbaum-Kadanoff-Shenker,Siggia-Rand}. For this last case, frequency lockings with devil's staircase \cite{CircleMap} and doubling or fractalization of tori \cite{KK-ProgTheor83} are uncovered in regard to two-dimensional tori.

Transition from a torus with more than two frequencies to chaos, in contrast, is not fully explored, in spite of its historical significance in the investigation of fluid turbulence. 
Landau proposed to understand turbulence as (possibly an infinite number of) successive Hopf bifurcations leading to a torus with an infinite number of (incommensurate) frequencies \cite{Landau-Lifshitz}. In contrast, it was noted that at any vicinity of a dynamical system having a three-dimensional-torus attractor, there can appear chaotic motion following some structural perturbations to the dynamical system \cite{Ruelle-Takens,Newhouse-Ruelle-Takens,Turaev2015}. This theorem, on the one hand, promoted the experimental studies on the onset of turbulence by low-dimensional chaos \cite{Berge-Pomeau-Vidal,Experiment34}; on the other hand, it was sometimes overestimated and misinterpreted as tori of higher than two dimensions being easily destabilized and replaced by chaos with most slight structural modifications  (i.e., by adding any nonlinear term). The latter, indeed, is not the case. In fact, the mathematical theorem does not say anything about how large the fraction of chaos attractors is in the vicinity of high-dimensional tori.

With subsequent numerical and experimental studies, the existence of three- or four-dimensional torus attractors has been confirmed \cite{Experiment34,KK1984,Grebogi-Ott-Yorke,Kim-Mackay}, where structural perturbations to such dynamical systems often lead to lower-dimensional tori due to frequency lockings, rather than chaotic attractors. However, of note is that 
chaos can appear even for weak nonlinearity, albeit in a small fraction. This is in contrast to the case with two-dimensional tori for which sufficiently strong nonlinearity is necessary for chaos to appear. Still, the nature of chaos near high-dimensional tori as well as the transition to it has not been well explored. 

In this Letter, we explore this long-lasting problem of transition to chaos from high-dimensional torus attractors. We mainly address three questions. The first question is with regard to the existence of high-dimensional tori---Can they still exist as attractors with the increase in dimension of tori? We will present how the fractions of tori and chaos depend on the dimensionality and nonlinearity of the system. 

The second question is with regard to the nature of chaos that emerges from a high-dimensional torus. We uncover the existence of chaos on (or in the vicinity of) a high-dimensional torus, as characterized by a large number of null (or close to null) Lyapunov exponents, accompanied by only one or a few positive exponents. The eigenvector for the maximum positive exponent is extended over degrees of freedom and shows slow itinerant motion with the $1/f^\nu$ spectrum, suggesting characteristics of chaos on a high-dimensional torus.

The third question to be addressed is the nature of the transition from tori to chaos, and the possible relevance of fractal torus to such transition. 

\begin{figure*}[htb]
\centering \includegraphics[width = 16 cm]{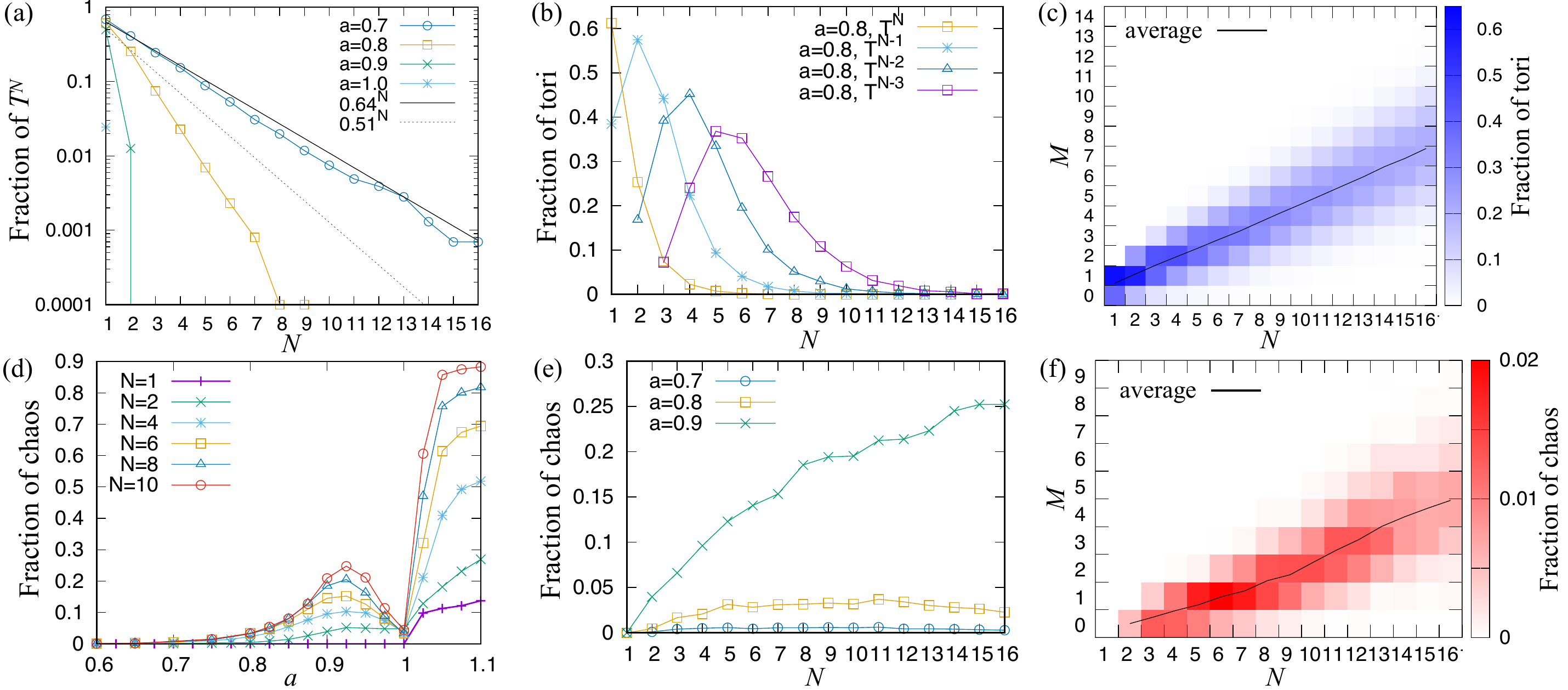} 
\caption{Fractions of tori and chaos in map (1) over a random choice of natural frequencies $\vec{\Omega}$. The coupling constant $J_{ij}=b$ is set to $0.1$, while $10^4$ samples of $\vec\Omega$ are randomly chosen as $\Omega_i\in \{0.0005,0.001,\cdots,0.5\}$. 
(a) Dependence of the fraction of $T^{N}$ on $N$. 
(b) Dependence of the fractions of $T^N$, $T^{N-1}$, $T^{N-2}$, and $T^{N-3}$ on $N$. 
(c) Fractions of $M$-dimensional tori for $a=0.8$. 
(d) Dependence of the fraction of chaos on $a$ and $N$. The critical value $a_c$ to lose invertibility is $1$ for $N=1$, and $>0.9$ for $N\geq2$: e.g., $a_c(2,0.1)\simeq 0.91$ and $a_c(4,0.1)\simeq 0.94$. 
(e) Dependence of the fraction of chaos on $N$. 
(f) Fractions of $M$-dimensional toric chaos for $a=0.8$. 
\label{fig:N-Torus_report2}}
\end{figure*}

To be specific, we study the following coupled circle map: 
\begin{eqnarray}x_{i}(t+1)=x_i(t)+\Omega_i+\frac{a}{2\pi}\Bigl[\sin(2\pi x_i(t))\nonumber\\+\frac{1}{N-1}\sum_{j=1;j\neq i}^N J_{ij}f\left(x_i(t),x_j(t)\right)\Bigr],
\end{eqnarray}
where $N$ oscillators are heterogeneous (i.e., have different natural frequencies $\Omega_i$) and globally coupled \cite{N=1}. Here, we mainly consider the model with $J_{ij}=b$ and $f\left(x_i(t),x_j(t)\right)=\sin(2\pi x_j(t))$, although the results below do not depend on the precise form of them \cite{Settings}.

The map is an extension of globally coupled map to include heterogeneity over elements, and a discrete-time version of coupled oscillators \cite{Kuramoto,KK-GCM}. It can be generally regarded as a map on the oscillation phases for Poincar\'e map of a flow obeying some ordinary differential equation. Hence, the attractors with $M$ incommensurate frequencies (i.e., with $M$ null Lyapunov exponents, without positive ones) correspond to $(M+1)$-dimensional tori in the continuous-time flow system. Another important feature of map (1) is that when $a< {}^\exists a_c(N,b)$, the Jacobian is always non-zero, and thus the dynamical system is invertible \cite{Invertible}. For example, $a(1+b)<1$ is a sufficient condition for the invertibility in the case of $J_{ij}=b$. 

The case with $N=1$ was first introduced as a (sine)-circle map by Arnold for the study of two-dimensional tori, i.e., as a map on one phase of oscillation sampled by the inverse of the other frequency. In this case, the map is shown to be invertible for $a<a_c(1,b)=1$ where the attractor is either a torus or a periodic cycle; whereas, for $a>1$, it is either chaos or a periodic cycle. 
This map has been investigated as a standard model for the transition to chaos (at $a=a_c$) from quasiperiodicity on a two-dimensional torus \cite{Feigenbaum-Kadanoff-Shenker,Siggia-Rand,CircleMap,KK-ProgTheor83}. 
In contrast, for $N = 2$, which corresponds to the map for a three-dimensional flow, chaos can appear even for $a<a_c(2,b)$ \cite{KK1984,Grebogi-Ott-Yorke,Kim-Mackay}. 

To examine the nature of each attractor, we computed the $N$-dimensional Lyapunov spectra by iterating map (1) $2.5\times10^6$ times after dropping the initial $5\times10^5$ times of iterations \cite{zero}. 

\begin{figure*}[htb]
\centering \includegraphics[width = 17.8 cm]{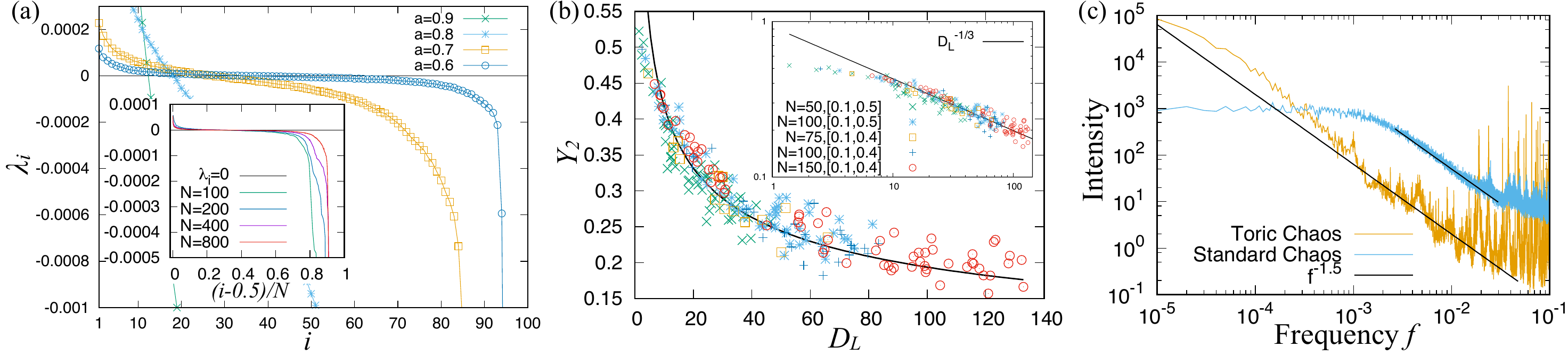} 
\caption{Properties of toric chaos. 
(a) Lyapunove spectra $\vec{\lambda}$ with $N=100$ and $b=0.7$. Different colors correspond to different nonlinearity $a$, while natural frequencies $\vec\Omega$ are uniformly distributed in $(0.1,0.5]$. (inset) the scaled plot with $a=0.65$ and $b=0.5$: $\lambda_i$ vs $(i-0.5)/N$ for $N=100,200,400,800$. (b) The inverse participation ratio of the first Lyapunov vector $Y_2$ vs Lyapunov dimension $D_L$. $Y_2$ is calculated as the average over $2.5\times 10^6$ steps across different parameters of $a=0.53\sim0.98$, $b=0.1\sim0.9$, within the invertible regime for $N=50,75,100,150$. $\vec\Omega$ are uniformly distributed in $(0.1,0.5]$ or $(0.1,0.4]$. The black line shows the curve $Y_2\propto D_L^{-1/3}$. (inset) the log--log plot.
(c) Power spectra of the first Lyapunov vector for toric chaos (orange; $a=0.65,b=0.7$) and standard chaos (blue; $a=0.98,b=0.5$) with $N=100$. 
The slopes of the black lines are $-1.5$. 
\label{fig:LyapunovVec-PowerSpectrum}}
\end{figure*}

First, the fractions of tori and chaos were numerically computed, by randomly choosing the natural frequencies $\vec{\Omega}$. 
As shown in Fig. \ref{fig:N-Torus_report2}(a), even the highest-dimensional (i.e., $N$-dimensional in map or $(N+1)$-dimensional in flow) tori exist, while their fraction decreases exponentially with $N$. 
This decrease is mostly due to locking to lower-dimensional tori, rather than due to replacement by chaos [Figs. \ref{fig:N-Torus_report2}(b)-(c)]. 
As a result, the average dimensionality of torus attractors linearly increases with $N$ and its saturation with the increase in $N$ is not observed [Fig. \ref{fig:N-Torus_report2}(c)]. 
It suggests that high-dimensional (i.e., $O(N)$-dimensional) tori considerably exist, even for $N\geq3$. 

The exponential decrease in the highest-dimensional tori against $N$ can be understood as follows. Consider a dynamical system consisting of $N$ modes with incommensurate frequencies. If the modes do not (or little) interact with each other, an $N$-dimensional torus will exist as an attractor. Now, consider the addition of another mode that has a certain interaction with the other modes: the ratio of its frequency with those of the pre-existing modes may be close to a rational value, which leads to locking to a lower-dimensional torus. As a simple rough estimate, let us assume that such locking occurs with a certain probability, $p$. Then, the probability that no locking occurs is estimated as $(1-p)^N$. 
This gives a rudimentary understanding of why the fraction of $N$-dimensional tori decreases exponentially \cite{ExpoDecrease}. 

In the case with $N\geq 2$, chaos can appear even in the invertible regime, in contrast to the case with $N=1$ in which chaos can exist only in the non-invertible regime $a>a_c$ [Figs. \ref{fig:N-Torus_report2}(d)-(f)]. Following Ruelle--Takens--Newhouse \cite{Ruelle-Takens,Newhouse-Ruelle-Takens} (and earlier studies cited in Ref. \cite{Ruelle-Takens}), chaos can appear from a (high-dimensional) torus even in the weak nonlinearity (i.e., invertible) regime, $a<a_c$. Here, in the non-invertible regime, the fraction of complete locking to fixed points (rather than chaos) increases near $a=1$, as in the case with $N=1$ [see Fig. \ref{fig:N-Torus_report2}(d) and Supplemental Material \cite{SupplementalMaterial}, Fig. S1]. Figures \ref{fig:N-Torus_report2}(d)-(e) reveal that the fraction of chaos increases with $N$, while the increase is saturated in the invertible regime. 

Remarkably, chaos in the invertible regime often has not only positive Lyapunov exponent(s) but also multiple Lyapunov exponents that are exactly or nearly $0$ [Fig. \ref{fig:N-Torus_report2}(f)], in contrast to that in the non-invertible regime. In other words, such chaos exists on (or in the vicinity of) a torus. Hence, we term this form of chaos with $M$ null Lyapunov exponents as $M$-dimensional ``toric chaos'' (or, $(M+1)$-dimensional in flow). 

Note that, in the invertible regime, $\sum_{i=1}^N\lambda_i\leq0$ always holds, as is also confirmed numerically. 
Thus, toric chaos cannot appear for two-dimensional maps (which correspond to three-dimensional flows studied earlier following the line of Ruelle--Takens scenario). 
The appearrance of toric chaos is possible only for the map with $N\geq 3$ (i.e., $\geq 4$-dimensional flows) that allows for the Lyapunov spectrum $(+,+,\cdots, 0,0,\cdots, -,-,\cdots)$ [see Fig. \ref{fig:N-Torus_report2}(f)]. 
Then, how many null Lyapunov exponents can toric chaos have in a high-dimensional system? To address this question, we studied map (1) by increasing $N$ to $\geq50$, where the natural frequencies $\vec{\Omega}$ are uniformly distributed in an interval. 
Figure \ref{fig:LyapunovVec-PowerSpectrum}(a) shows the Lyapunov spectra $\vec{\lambda}$ for different $N$ plotted by scaling the index as $(i-0.5)/N$. 
It suggests that the torus dimension, $M$, is an extensive variable, proportional to $N$. 
Indeed, this is consistent with the result in Fig. \ref{fig:N-Torus_report2}(f) that the average number of neutral Lyapunov modes in toric chaos increases linearly with $N$. Toric chaos shows the accumulation of $O(N)$ null Lyapunov exponents.

\begin{figure*}[htb]
\centering \includegraphics[width = 17.8 cm]{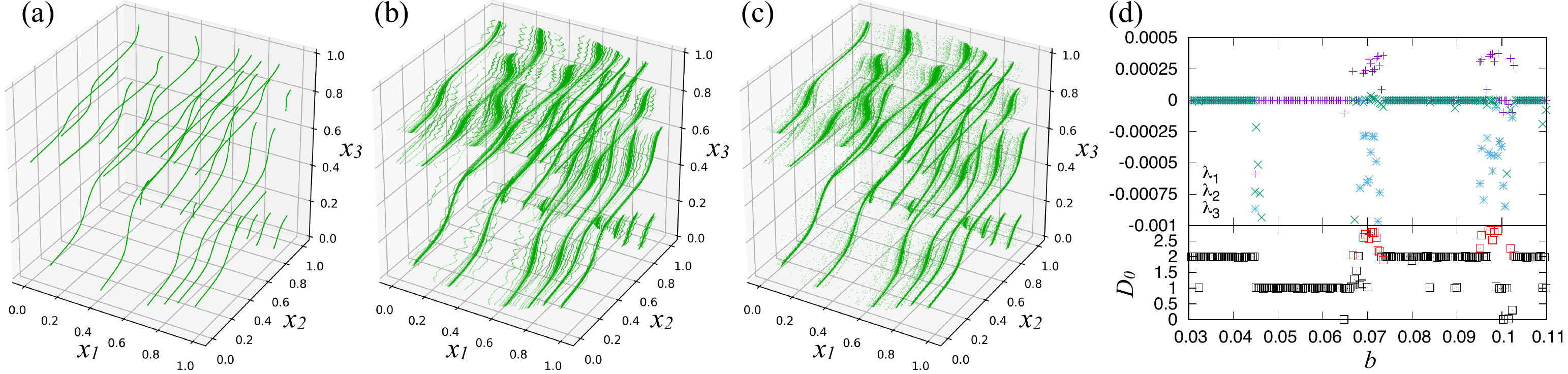}
\caption{
Bifurcation in map (1) with the change of coupling constant $b$ for $N=3,a=0.7,\vec{\Omega}=(0.198,0.4795,0.1945)$. 
(a-c) $(x_1(t),x_2(t),x_3(t))$ are plotted for $2\times 10^5< t\leq 4\times 10^5$. 
(a) $b=0.06$ ($1$-torus). $\vec\lambda=(5\times10^{-9},-0.002,-0.01)$, $D_0\simeq 1.0$, and $D_L=1$. 
(b) $b=0.06675$ (fractal $1$-torus). $\vec\lambda=(1\times10^{-8},-0.0002,-0.002)$, $D_0\simeq 1.5$, and $D_L=1$. 
(c) $b=0.0667$ (chaos). $\vec\lambda=(0.0002,4\times10^{-7},-0.002)$, $D_0\simeq 2.05$, and $D_L\simeq 2.1$. 
(d) Dependence of Lyapunov exponents $\lambda_i$ (top) and box counting dimension $D_0$ (bottom) on $b$. The box counting dimensions for chaos (torus), computed for $240^3$ bins, are shown in red (black) symbols. Clear deviations of $D_0$ from $0$, $1$, or $2$ for black symbols correspond to fractal-torus attractors.
\label{fig:FractalizationOfTorus}}
\end{figure*}

Now, the characteristic behaviors of $M$-dimensional toric chaos will be studied. To examine whether chaotic motion spreads over $M$-dimensional tori or is localized, we computed the behavior of the first Lyapunov vector $\vec{v}^{(1)}$. 
First, the localization of the ``chaos mode'' is quantified by its inverse participation ratio, $Y_2\equiv\braket{\sum_{i=1}^N|v_i^{(1)}|^4}_t$; if $\vec{v}^{(1)}$ is extended to all $N$ elements, $Y_2$ is $O(1/N)$, and if the vector is localized, $Y_2$ is $O(1)$ \cite{Mirlin2000,KK-PhysicaD86}. 
Figure \ref{fig:LyapunovVec-PowerSpectrum}(b) shows the negative correlation between $Y_2$ and the Lyapunov dimension $D_L$, where $Y_2$ is approximately proportional to $1/D_L^\alpha$ with the exponent $\alpha\approx 1/3$, independent of the parameters $N,\vec{\Omega},a,b$. 
Here, note that the Lyapunov dimension of $M$-dimensional toric chaos is at least $M+1$, and thus $D_L$ approximately quantifies the torus dimension $M$ of toric chaos in the invertible regime \cite{Non-InvertibleY2}. 
For standard chaos in high-dimensional dynamical systems, $Y_2$ is $O(1)$ against the increase in the dimension $N$, i.e., the Lyapunov vector for the chaos mode is localized. 
In contrast, the decrease of $Y_2$ for toric chaos against $M$ (and $D_L$) implies that the chaos mode is spread over a high-dimensional torus. Still, it is noted that $\alpha\approx1/3$ is less than $1$, which suggests partial localization compared with the fully extended case. 

To see the nature of this localization, the temporal event of $\vec{v}^{(1)}$ is plotted in Supplemental Material \cite{SupplementalMaterial}, Fig. S3. This suggests partial localization of the vector, as well as its slow itinerant motion. We then computed the power spectra of the first Lyapunov vector $\vec{v}^{(1)}$. As shown in Fig. \ref{fig:LyapunovVec-PowerSpectrum}(c), $1/f^\nu$ fluctuation is observed for toric chaos, with the exponent $\nu\approx 3/2$, whereas the standard chaos does not exhibit such $1/f^\nu$ spectra. As the torus dimension increases, $1/f^\nu$ behavior is extended to much lower frequencies.

Lastly, the transition from a torus to chaos is studied by using the case with $N=3$. In Figs. \ref{fig:FractalizationOfTorus}(a)-(c), examples of torus and toric-chaotic attractors are shown (see also Supplemental Material \cite{SupplementalMaterial}, Figs. S4 and S5 for other examples). 
As depicted in Fig. \ref{fig:FractalizationOfTorus}(d), the dimensionality (i.e., box counting dimension $D_0$) of a torus first decreases with the increase in the coupling constant $b$, and at some value of $b$, chaos appears ``on the torus.'' In the vicinity of the transition to toric chaos, the torus exhibits oscillation down to a smaller wavelength, in other words, strange nonchaotic (or, fractal-torus) attractors with non-integer dimensions exist [see Fig. \ref{fig:FractalizationOfTorus}(b)] \cite{N=2}, whereas it is difficult to prove the fractality rigorously through numerical simulations. 
At least, data suggests the transition from a torus to toric chaos occurs via fractalization of the torus (i.e., strange nonchaotic attractor; SNA) \cite{Nishikawa,Feudel-Kuznetsov-Pikovsky,Grebogi-SNA},
although frequency lockings to lower-dimensional torus intervened in by parameter $b$ make it harder to confirm this.

In summary, we investigated globally coupled circle maps to understand if and how chaos appears from a high-dimensional torus. First, we found that $T^{N}$ exist even for large $N$. Next, despite the prevalence of torus attractors, chaos can appear even in the invertible regime (i.e., for lower nonlinearity compared to the case with $N=1$), and its fraction increases with $N$ and is then saturated. In case of $N\geq 3$, chaos in the invertible regime often exists ``on a torus.'' Such chaos, termed toric chaos, has a number ($O(N)$) of null Lyapunov exponents. This toric chaos has both the faces of torus and chaos: the chaos mode is extended on a torus (not localized as in standard chaos), but is not spread equally over all the dimension of the torus. It is characterized by the anomalous decrease of the inverse partition ratio of the first Lyapunov vector, $Y_2$, against the torus dimension $M$ (and Lyapunov dimension $D_L$). Slow itinerant motion of the chaos mode is also observed, suggesting slower changes in the stretching directions. 
 
Indeed, long-term, chaotic change over lower-dimensional states (termed as attractor ruins) is known as chaotic itinerancy, where the accumulation of Lyapunov exponents close to $0$ is reported \cite{ChaoticItinerancy,InformationCascade}, similar to the present toric chaos. In this respect, toric chaos might be interpreted as chaotic itinerancy over lower-dimensional tori on a high-dimensional torus. 

We also discussed the possible relevance of fractal torus to the transition to toric chaos. The existence of fractal torus was investigated as SNA in a forced system \cite{Feudel-Kuznetsov-Pikovsky,Grebogi-SNA,Nishikawa}, while no conclusive evidence for its existence is given in an autonomous system thus far. It is often believed that SNA exists not in autonomous systems but only in forced systems; though, in the present case with $N \ge 3$, one neutral mode remains for the transition from torus to toric chaos, and that mode might work as quasiperiodic forcing, which may allow for SNA at this transition. 

Note that the torus dimension is extensive, i.e., the number of null Lyapunov exponents, $M$, increases as $O(N)$ when $N\to\infty$. In addition, the above results are qualitatively reproduced even with different forms of coupling. These indicate the generality of the existence of high-dimensional tori and toric chaos on them. Accordingly, one may expect a variety of possible applications of toric chaos. Here, we briefly point out just two of them.
 
In electroencephalogram (EEG), several peaks with different frequencies exist in its power spectra---alpha, beta, gamma, delta, theta, and mu waves. EEG also involves continuous power spectra besides these peaks, suggesting the existence of rich aperiodic (chaotic) modes, together with quasiperiodicity with different frequencies. Moreover, possible relevance of chaotic itinerancy for autonomously switching internal modes is discussed both theoretically \cite{KK-GCM,ChaoticItinerancy,Tsuda} and experimentally \cite{LeslieKay,LeslieKay2003,Freeman}. Although the complexity of neural dynamics is much higher, toric chaos may provide a novel perspective to it. 
 
Finally, we come back to the original problem in this Letter, concerning turbulence. As mentioned already, the relevance of high-dimensional tori to turbulence has been discussed in many works since Landau \cite{Landau-Lifshitz} (see also Ref. \cite{Spain} for recent proposition); whereas the viewpoint on low-dimensional chaos has been dominant after the proposition by Ruelle and Takens. 
Probably, neither of the two pictures is sufficient to understand the full spectrum of turbulence. The toric chaos, in this respect, may provide a fresh view to the old problem, as it captures both high-dimensional torus with a broad range of frequencies, and chaotic dynamics needed for turbulence.

\setcounter{figure}{0}
\renewcommand{\thefigure}{S\arabic{figure}}
\renewcommand{\thesection}{S\arabic{section}}

\hrulefill
\section*{Supplemental Material}

\begin{figure*}[hbt]
\centering \includegraphics[width = 15 cm]{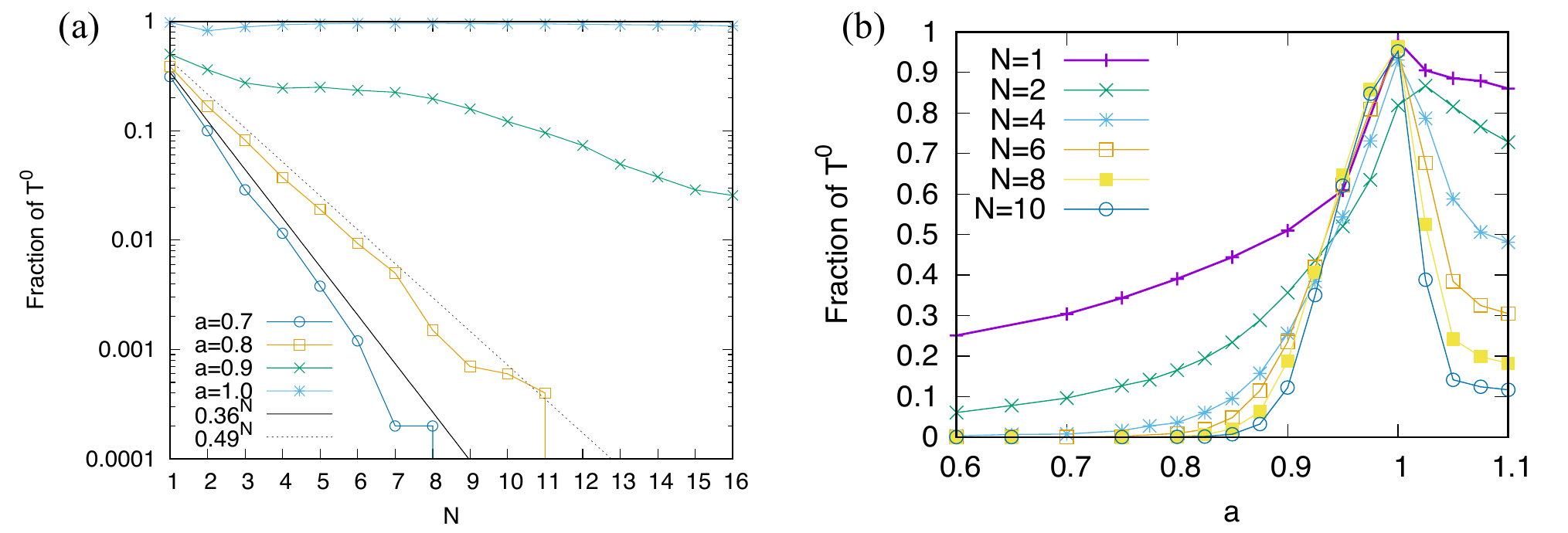}
\caption{The fractions of complete locking, $T^0$, in map (1) in the main text. 
(a) Dependence of the fraction of $T^0$ on $N$. 
(b) Dependence of the fraction of $T^0$ on $a$. \label{fig:Omake}}
\end{figure*}

\begin{figure*}[hbt]
\centering \includegraphics[width = 15 cm]{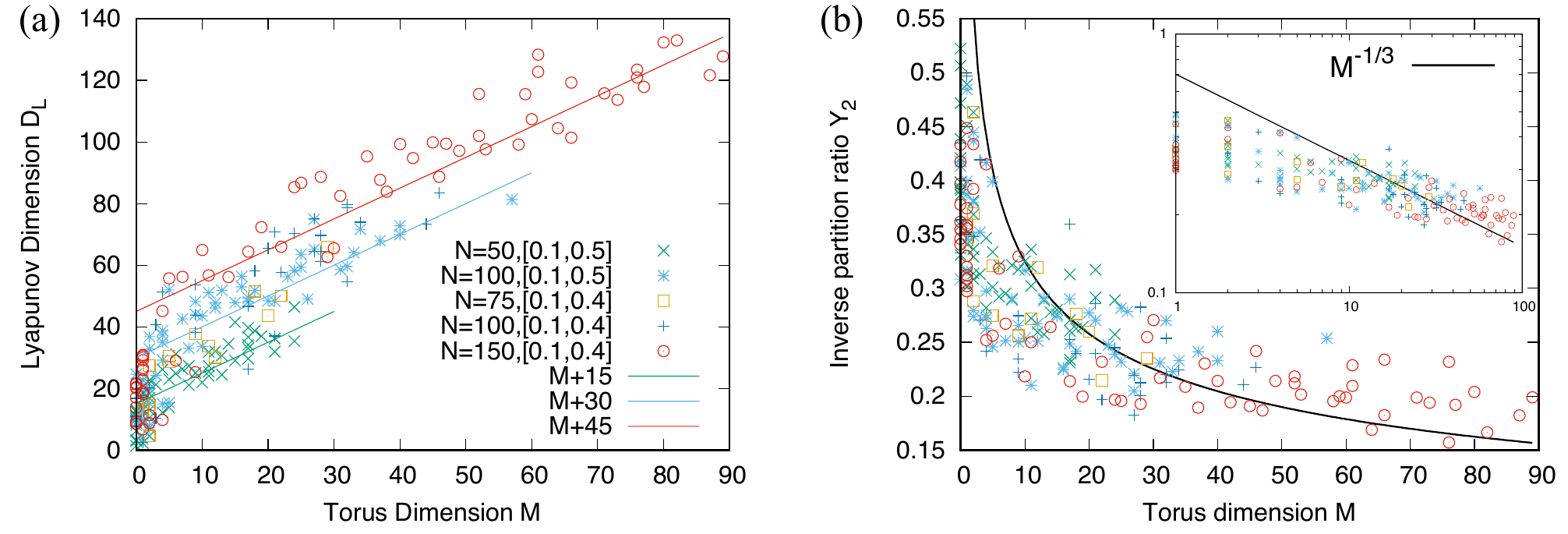}
\caption{Relationship among torus dimension $M$, Lyapunov dimension $D_L$, and $Y_2$. 
(a) $D_L$ vs $M$. \textcolor{black}{Approximately, $D_{\rm L}\simeq M+0.3N$ holds.} 
(b) $Y_2$ vs $M$. \label{fig:Omake2}}
\end{figure*}

\begin{figure*}[hbt]
\centering \includegraphics[width = 15 cm]{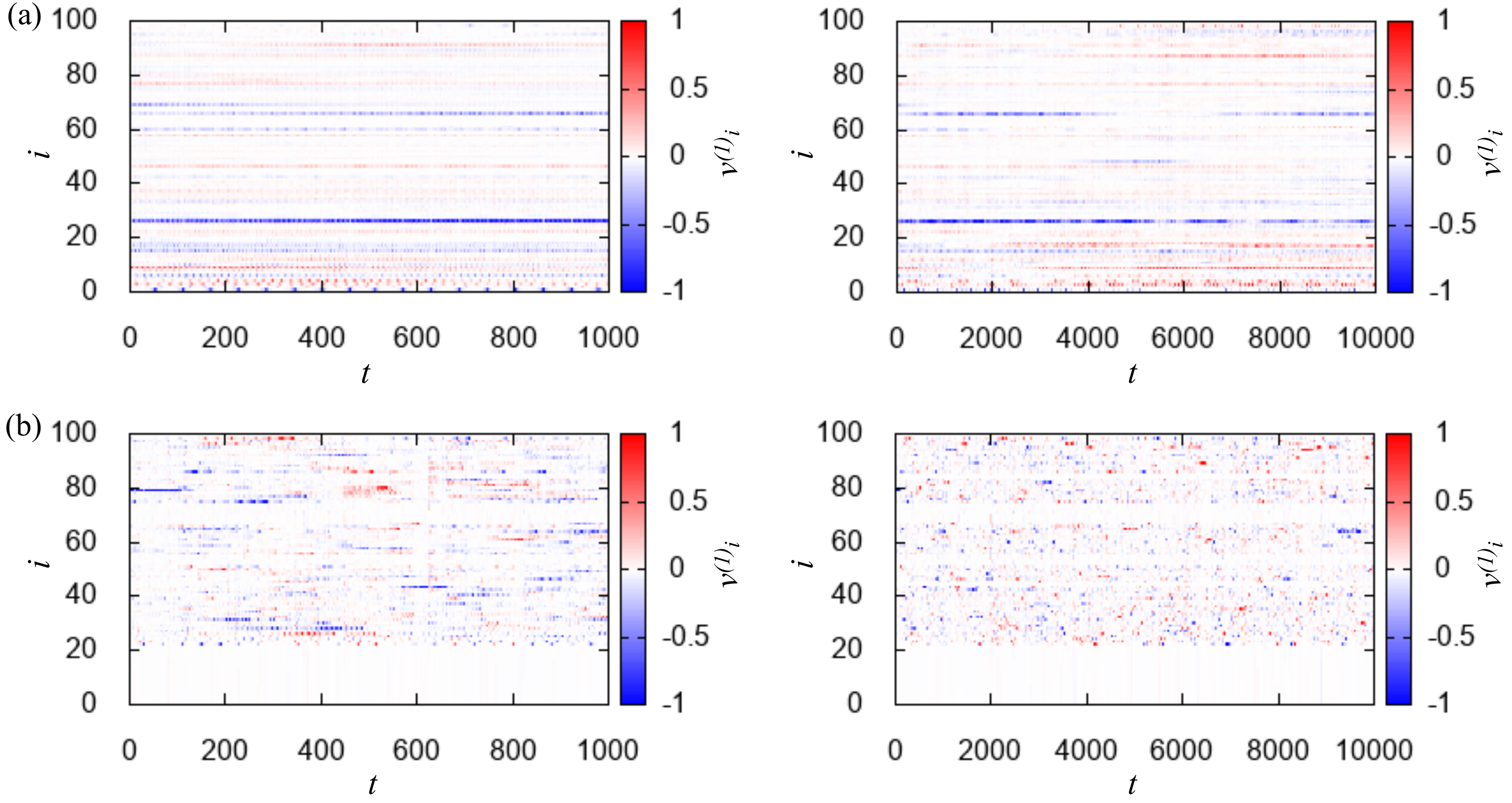} 
\caption{The time series of the first Lyapunov vector, $\vec{v}^{(1)}(t)$, in map (1) in the main text are plotted for every step (left) and every ten steps (right). 
$\vec\Omega$ are uniformly distributed in $(0.1,0.5]$, and the parameters are set as $N=100$ and $b=0.7$. 
(a) Toric chaos with $a=0.6$. (b) Standard chaos with $a=0.98$. \label{fig:50-Dim}}
\end{figure*}

\begin{figure*}[hbt]\centering \includegraphics[width = 16 cm]{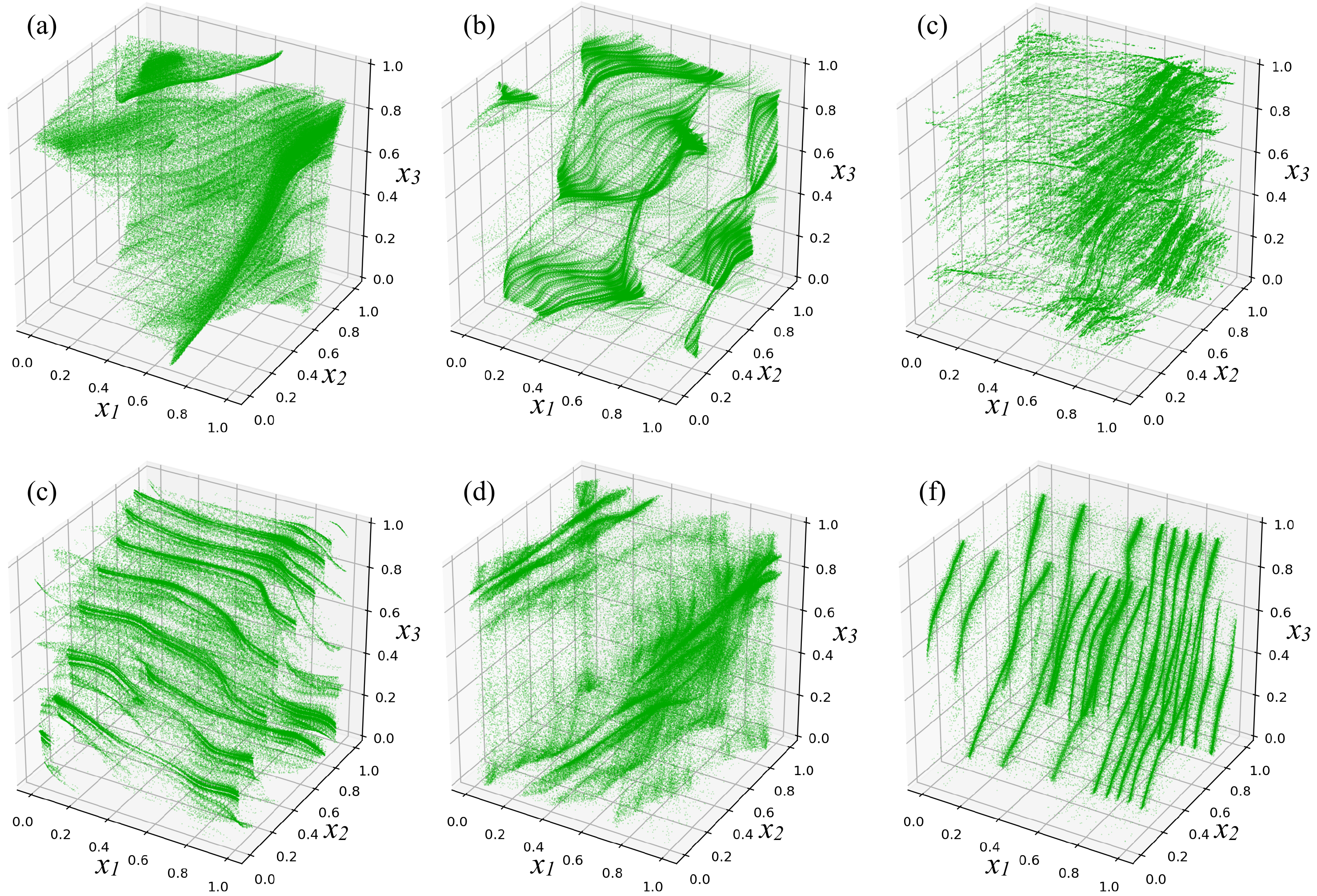}
\caption{Examples of toric-chaos attractors with various natural frequencies $\vec{\Omega}$. 
$(x_1(t),x_2(t),x_3(t))$ are plotted for $2\times 10^5< t\leq 4\times 10^5$. 
(a) $a=0.7, b=0.0666, \vec{\Omega}=(0.195,0.1965,0.1975)$. $\vec\lambda=(0.001,-6\times 10^{-7},-0.004)$. 
(b) $a=0.7, b= 0.1, \vec{\Omega}=(0.235,0.2815,0.4715)$. $\vec\lambda=(0.00015,-6\times 10^{-7},-0.0002)$. 
(c) $a=0.7, b= 0.1, \vec{\Omega}=(0.1305,0.277,0.2745)$. $\vec\lambda=(0.0005,2\times 10^{-6},-0.01)$. 
(d) $a=0.8, b= 0.0679, \vec{\Omega}=(0.3205,0.3175,0.3155)$. $\vec\lambda=(0.0006, -2\times 10^{-8},-0.003)$. 
(e) $a=0.8, b= 0.1, \vec{\Omega}=(0.2625,0.466,0.2655)$. $\vec\lambda=(0.0003,-9\times 10^{-7},-0.0004)$. 
(f) $a=0.8, b= 0.1, \vec{\Omega}=(0.1965,0.2475,0.439)$. $\vec\lambda=(0.0002,-6\times 10^{-7},-0.0002)$. 
\label{fig:ToricChaosExs}}
\end{figure*}

\begin{figure*}[hbt]\centering \includegraphics[width = 16 cm]{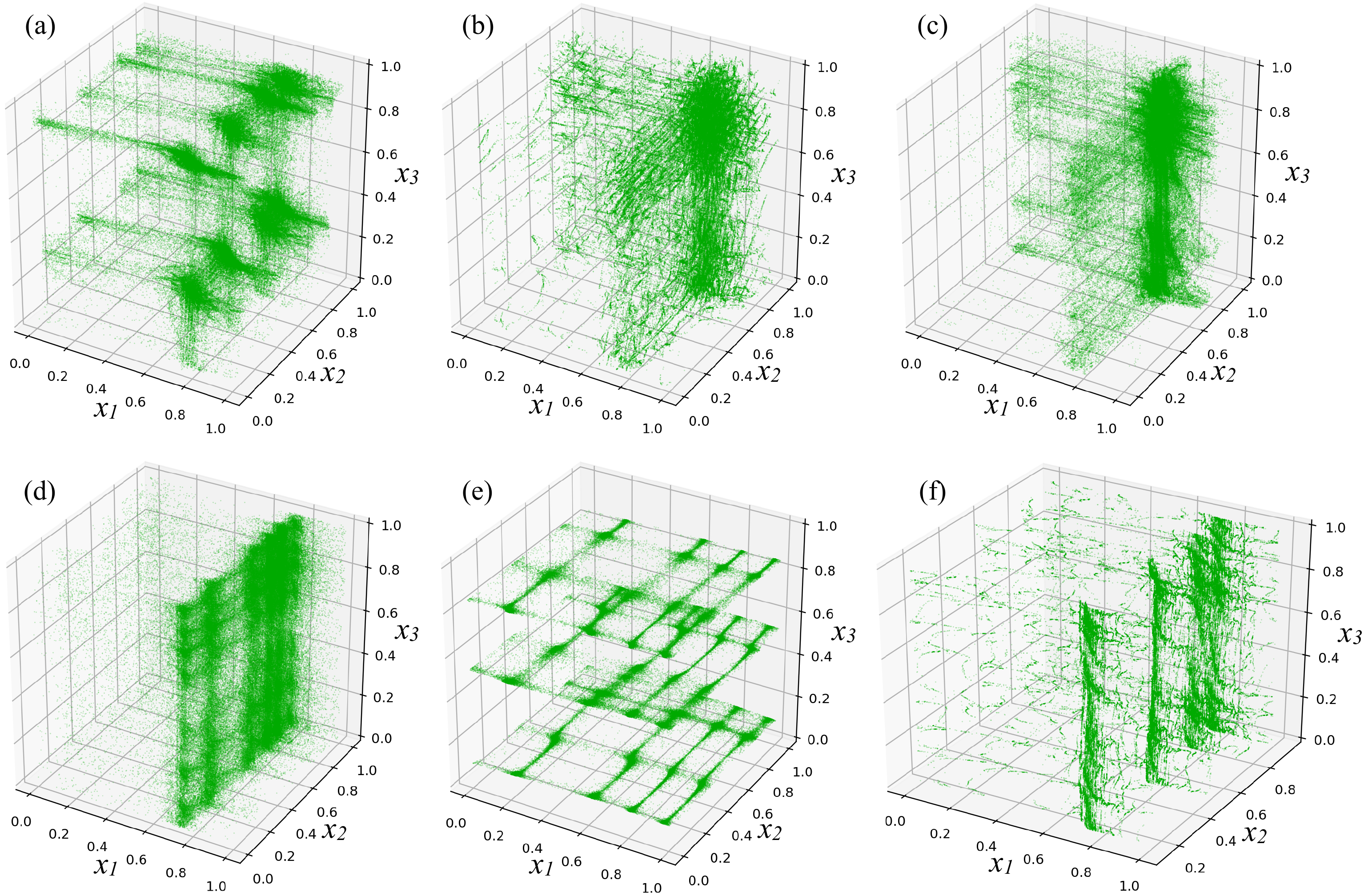}
\caption{Examples of standard-chaos attractors (i.e., non-toric chaos) with various natural frequencies $\vec{\Omega}$. 
$(x_1(t),x_2(t),x_3(t))$ are plotted for $2\times 10^5< t\leq 5\times 10^5$. 
(a) $a=0.7,b=0.1,\vec{\Omega}=(0.1155,0.2695,0.486)$. $\vec\lambda=(0.004, -0.0005 -0.008)$. 
(b) $a=0.8,b=0.0667,\vec{\Omega}=(0.1365,0.151,0.195)$. $\vec\lambda=(0.0005,-0.001,-0.004)$. 
(c) $a=0.8,b=0.0667,\vec{\Omega}=(0.134,0.137,0.214)$. $\vec\lambda=(0.001,-0.0004,-0.003)$. 
(d) $a=0.8,b=0.0667,\vec{\Omega}=(0.1295,0.2385,0.3075)$. $\vec\lambda=(0.002,-0.0001,-0.004)$. 
(e) $a=0.8,b=0.0667,\vec{\Omega}=(0.2775,0.3385,0.35)$. $\vec\lambda=(0.001,-0.001,-0.2)$. 
(f) $a=0.9,b=0.1,\vec{\Omega}=(0.147,0.287,0.391)$. $\vec\lambda=(0.0002,-0.001,-0.2)$. 
\label{fig:nonToricChaosExs}}\end{figure*}

\begin{figure}[hbt]
\centering \includegraphics[width = 8 cm]{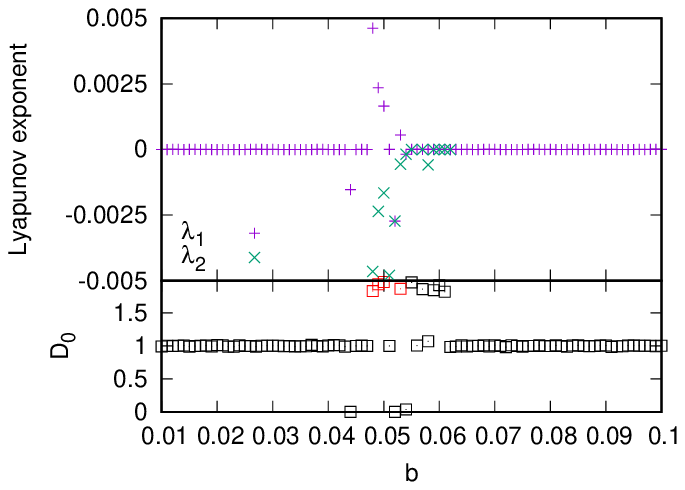}
\caption{Bifurcation diagram with $N=2$, $a=0.7$, and $\vec{\Omega}=(0.1125,0.271)$. Plotted in the same manner with Fig. 3(d) in the main text. 
\label{fig:N=2}}
\end{figure}


\section{Toric chaos commonly appears in various forms of coupling}
The results in the main text are qualitatively reproduced even with different forms of couplings, $J_{ij}f\left(x_i(t),x_j(t)\right)$. 

\begin{figure*}[hbt]
\centering \includegraphics[width = 15 cm]{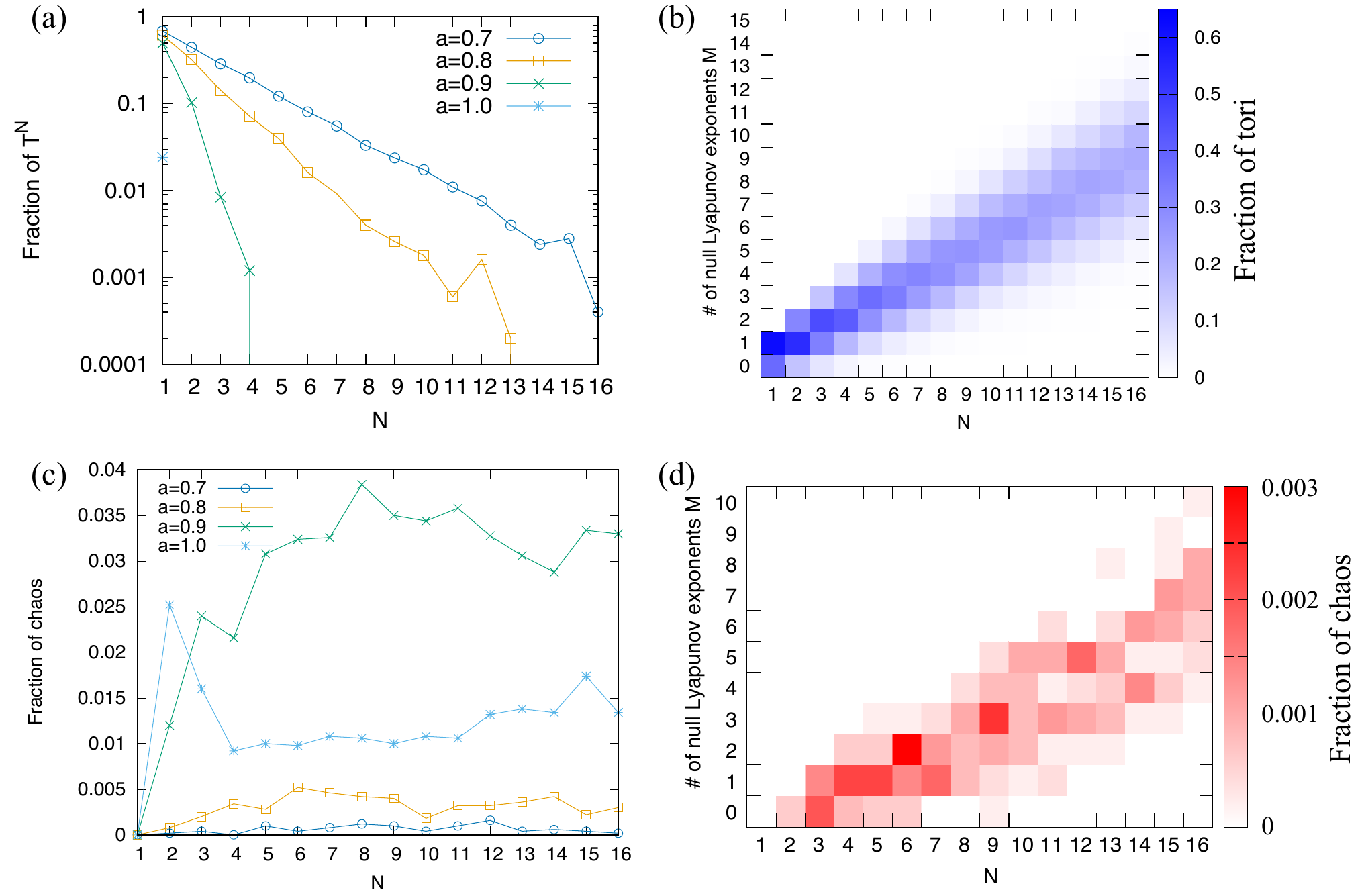} 
\caption{The coupling constants $J_{ij}$ are heterogeneous and uniformly-randomly chosen in $[-b,b]$. $T=10^6,b=0.1$. 
(a) Dependence of the fraction of $T^{N}$ on $N$. 
(b) Fractions of $T^M$ with $a=0.8$. 
(c) Dependence of the fraction of chaos on $N$. 
(d) Fractions of $M$-dimensional toric chaos with $a=0.8$. 
\label{fig:Torus-Chaos-NonUniform}}
\end{figure*}

(i) When the coupling constants $J_{ij}$ are heterogeneous and randomly chosen (e.g., $J_{ij}$ is chosen as a uniform random number in $[-b,b]$ or $[0,2b]$),  $N$-dimensional tori and toric chaos exists (Fig. \ref{fig:Torus-Chaos-NonUniform}), while the fraction of toric chaos decreases to some degree. 

\begin{figure*}[hbt]
\centering \includegraphics[width = 15 cm]{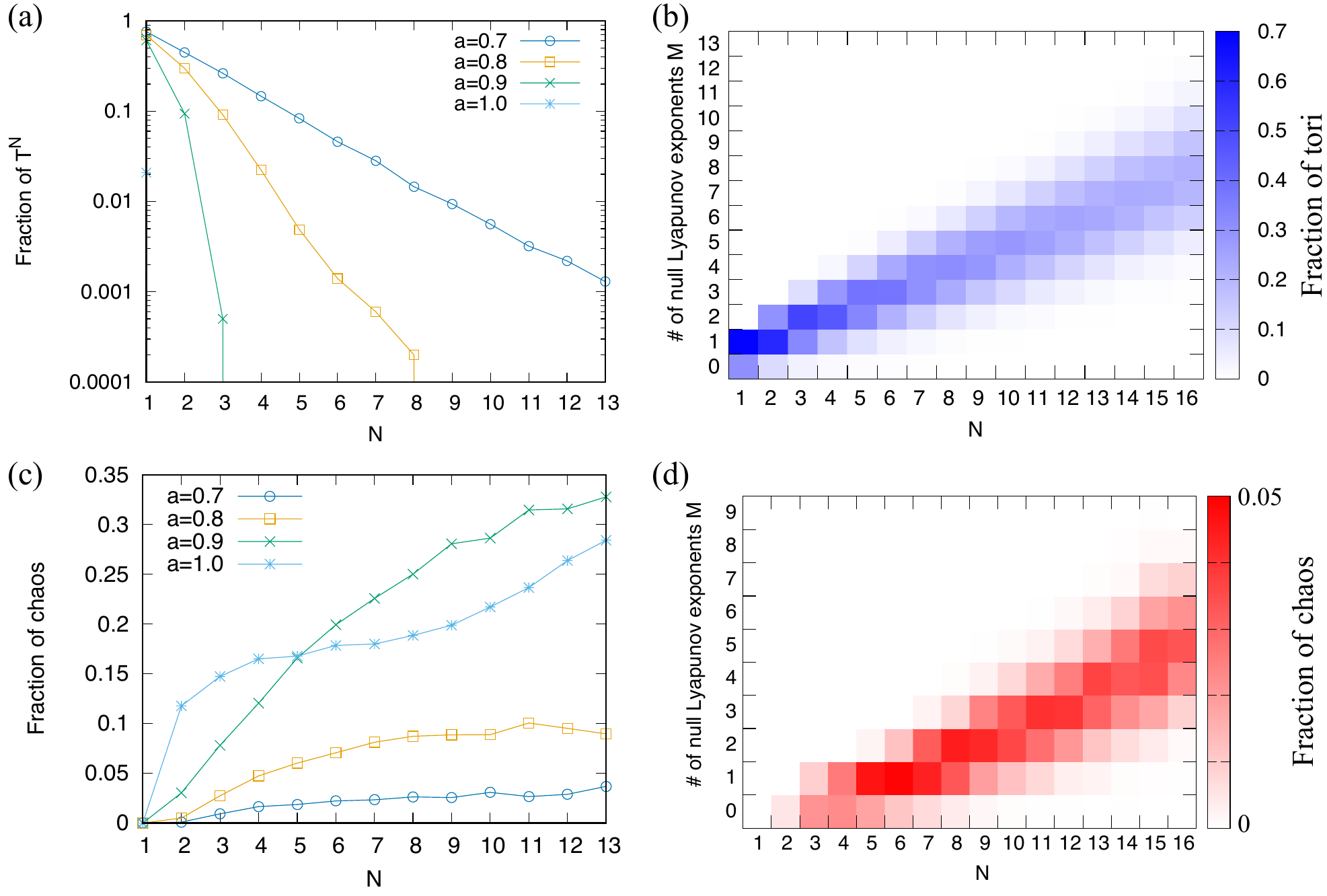} 
\caption{The case in which $J_{ij}=b$ is homogeneous and The coupling is given by $f\left(x_i(t),x_j(t)\right)= \sin\left(2\pi (x_i(t)-x_j(t))\right)$. $T=10^6,b=0.1$. 
(a) Dependence of the fraction of $T^{N}$ on $N$. 
(b) Fractions of $T^M$ with $a=0.7$. 
(c) Dependence of the fraction of chaos on $N$. 
(d) Fractions of $M$-dimensional toric chaos with $a=0.7$. 
\label{fig:Kuramoto}}
\end{figure*}

(ii) When the coupling term $f$ is Kuramoto-model-like, i.e., $f\left(x_i(t),x_j(t)\right)= \sin\left(2\pi (x_i(t)-x_j(t))\right)$, toric chaos and $N$-dimensional tori exists (Fig. \ref{fig:Kuramoto}).\\

\begin{thebibliography}{10}
\bibitem{N=1}
The last term in Eq. (1) is regarded as $0$ with $N=1$.
\bibitem{Invertible} 
For $N=2,3,4$, $a_c$ is calculated as $a_c(2,b)=1/(1+b)$, $a_{c}(3,b)=(\sqrt{4+4b^\prime-7{b^\prime}^2}-b^\prime)/(2+2b^\prime-4{b^\prime}^2)$, and $a_{c}(4,b)=(\sqrt{1+2b^\prime-2{b^\prime}^2}-b^\prime)/(1+2b^\prime-3{b^\prime}^2)$, with $b^\prime\equiv b/(N-1)$.
\bibitem{zero}
The Lyapunov exponent is regarded as zero if its value is less than $10^{-4}$ and larger than $-1\times 10^{-5}$.
\end{thebibliography}
\end{document}